\documentclass{aa}
\usepackage{times}
\usepackage{natbib}
\usepackage[dvips]{graphicx}
\bibpunct{(}{)}{;}{a}{}{,}

\def\pdot {\dot P}

\def\ltsima{$\; \buildrel < \over \sim \;$}
\def\lsim{\lower.5ex\hbox{\ltsima}}
\def\gtsima{$\; \buildrel > \over \sim \;$}
\def\gsim{\lower.5ex\hbox{\gtsima}}
\def\msun{~M_{\odot}}

\def\XMM{{\em XMM--Newton}}

\def\saxj{{SAX J0635+0533}}
\def\EPIC{{\em EPIC}}
\def\MOS{{\em MOS}}
\def\pn{{\em pn}}
\def\SAS{{\em SAS}}
\def\SSC{{\em SSC}}

\begin{document}

\title
{A low luminosity state in the massive X-ray  binary \saxj\ }

\author{Sandro Mereghetti\inst{1}, Nicola La Palombara\inst{1} }

\institute {INAF, Istituto di Astrofisica Spaziale e Fisica
Cosmica Milano, via E.\ Bassini 15, I-20133 Milano, Italy }

\offprints{S. Mereghetti, sandro@iasf-milano.inaf.it}

\date{Received: February 25, 2009 / Accepted: June 2, 2009}

\authorrunning{S. Mereghetti \& N. La Palombara}

\titlerunning{X-ray pulsar \saxj\ }

\abstract{The X--ray pulsar \saxj\ was repeatedly observed with
the XMM-Newton satellite in 2003-2004. The precise localization
provided by these observations confirms the association of \saxj\
with a Be star. The source was found, for the first time, in a low
intensity state, a factor $\sim$ 30 lower than that seen in all
previous observations. The spectrum, well fitted by an absorbed
power law with photon index $\sim$ 1.7 and N$_H$ =
1.2$\times$10$^{22}$ cm$^{-2}$, was compatible with that of the
high state. The low flux did not allow the detection of the
pulsations at 33.8 ms seen BeppoSAX and RXTE data. In view of the
small luminosity observed in 2003-2004, we reconsider the
peculiarities of this source in both the accretion and rotation
powered scenarios.

  }

\maketitle

\section{Introduction}

The X-ray source  \saxj\ was discovered with BeppoSAX in October
1997 \citep{kaa99} during a search for counterparts of
unidentified gamma-ray sources \citep{tho95}. Its 2-10 keV flux of
1.2 $\times$ 10$^{-11}$ erg cm$^{-2}$ s$^{-1}$, hard power law
spectrum (photon index $\sim$1.5) extending to 40 keV, and
positional coincidence with a V = 12.8 star of Be spectral type,
immediately suggested to classify \saxj\ as an accreting high mass
X-ray binary. The subsequent discovery of X-ray pulsations at 33.8
ms \citep{cus00} in the BeppoSAX data, makes this object quite
peculiar and raises some problems for the accretion scenario. In
fact, if the X-ray emission is powered by accretion on the neutron
star surface, in order to avoid the propeller effect (see, e.g.,
\citet{cam95}), the magnetic field in \saxj\ must be about three
orders of magnitude smaller than expected in a typical high mass
X-ray binary.

The pulse frequencies measured with RXTE in 1999, about two years
after the source discovery, indicated an orbital modulation with a
period of about 11 days and set a lower limit on the  long term
spin-down $\pdot$ $>$ 3.8$\times10^{-13}$ s s$^{-1}$ \citep{kaa00}.
Such a high $\pdot$ in a rapidly spinning neutron star implies  a
large rotational energy loss, ${\dot E}_{rot}$ = 10$^{45}$
4$\pi^{2}$ $\pdot$/$P^3$ = 4$\times$10$^{38}$ erg s$^{-1}$,
capable to power the observed X-ray luminosity without the need of
invoking mass accretion. In this interpretation, \saxj\ would
resemble other binary systems composed of a fast pulsar orbiting a
massive star, such as PSR B1259--63 \citep{jon92}, in which the
X-ray emission is thought to originate in the shock between the
pulsar's relativistic wind and that of the companion star. The
failure to detect radio pulsations from \saxj\ \citep{nic00}, does
not rule out this scenario, since beaming and/or absorption
effects might render the radio emission unobservable.

Here we report the results of  XMM-Newton observations, carried
out in 2003-2004, during which \saxj\ was detected at a very low
flux level, the smallest ever seen from this source.

\section{Observations and data analysis}
\label{obs}

\saxj\ was observed by \XMM\ with ten different pointings between
2003 September 11$^{th}$ and 2004 April 14$^{th}$. The three
\EPIC\ focal plane cameras \citep{tur01,str01} were active during
these pointings.  The two \MOS\ cameras were operated in the
standard {\em Full Frame} mode (time resolution 2.6 s), in order
to cover the whole 30\arcmin\ field--of--view.   The \pn\ camera
was operated in {\em Timing} mode, providing  a time resolution of
30 $\mu$s, but without imaging information.  For each observation
and focal plane camera the thin filter was used.

For each pointing we retrieved from the \XMM\ archive the
\textit{pps} files produced by \textit{pipeline processing
system}, which is operated by the \textit{Survey Science Center}.
Since only Timing mode data were obtained with  the \pn\ camera,
we concentrated on  the \MOS\ data to study the source properties.
Due to the faintness of  \saxj\ during these observations, it was
not possible to see the 33.8 ms pulsations in the \pn\ camera
data. For each observation, we looked for possible periods of high
instrumental background, caused by flares of soft protons with
energies less than a few hundred keV. To this aim, we selected
only single and double events (PATTERN $\leq$ 4) with energies
greater than 10 keV and recorded  in   the  peripheral CCDs. Then,
we  set  a count--rate  threshold  for good  time intervals  (GTI)
at 0.5  cts s$^{-1}$. By selecting only events within GTIs we
finally obtained a ``clean''  event list for  each \MOS\  data
set. The dates and effective exposure times (after soft--proton
rejection) of the observations are listed in
Table~\ref{detections}.

\begin{figure}[h]
\centering
\resizebox{\hsize}{!}{\includegraphics[angle=0,clip=true]{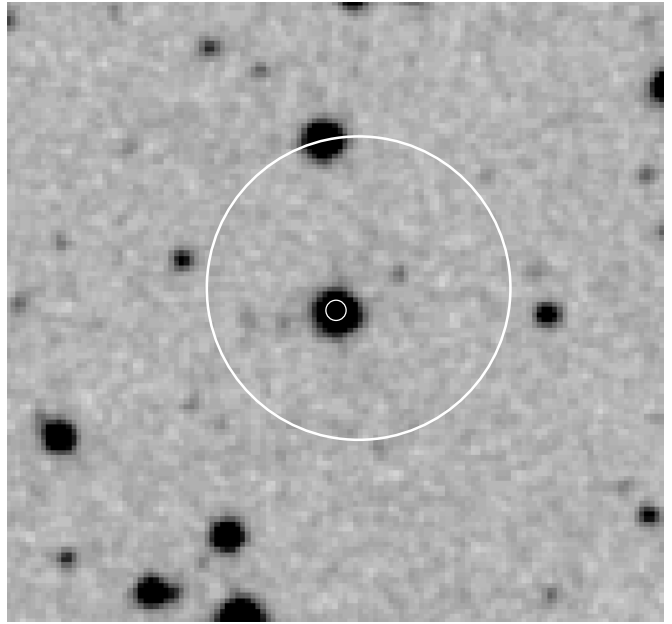}}
\caption{Optical image  of a 2$'$$\times$2$'$ field around the
position of \saxj\ from the Digitized Sky Survey (B filter). The
large circle is the BeppoSAX error region with a radius of 30$''$
\citep{cus00}. The small circle (2$''$ radius) is the \XMM\
position. North is to the top, East to the left. } \label{opt}
\end{figure}

After merging the event lists of the two \MOS\ cameras, we
accumulated an image of the field--of--view comprising all the
observations. This clearly showed a source, with a count rate of
(4.3 $\pm$ 0.3)$\times$10$^{-3}$ counts s$^{-1}$ in each camera, at
the coordinates R.A. = 06$^h$ 35$^m$ 18.3$^s$, Dec. = +05$^\circ$
33$'$ 06.3$''$ (J2000). This position was obtained after the
astrometric correction of the X--ray coordinates of the detected
sources, based on the positions of five optical counterparts found
in the Guide Star Catalogue \citep{Lasker2008}. The final
uncertainty on the source position is 1$''$ at 90\% c.l.
(including statistical and systematic errors). The \XMM\
localization confirms the association with the star proposed by
\citet{kaa99} on the basis of an estimated 4\% probability of
finding by chance a Be star in the 30$''$ radius error circle
determined with BeppoSAX (see Fig.~\ref{opt}).

We extracted the source spectra by selecting events in a circular
region with a small radius (10$''$)  in order to minimize the
background contribution. The   background spectra for the two
\MOS\ cameras were accumulated from large circular areas with no
sources and radii of 100$''$. We generated {\em ad hoc}
redistribution matrices and ancillary files using the \SAS\ tasks
\texttt{rmfgen} and \texttt{arfgen}, respectively. In order to
ensure the applicability of the $\chi^{2}$ statistics, all spectra
were rebinned with a minimum of 30 counts per bin, and fitted in
the energy range 0.3--10 keV using {\em XSPEC} 11.3.2. We checked
that separate fits of the two spectra gave consistent results,
therefore we analysed them simultaneously, in order to increase
the count statistics, imposing common spectral parameters for the
two spectra and introducing a free relative normalization factor
between them, to account for possible differences in the
instrument cross--calibration. We found a 1.02
cross--normalization factor between the \textit{MOS2} and the
\textit{MOS1} camera.

\begin{figure}[h]
\centering
\resizebox{\hsize}{!}{\includegraphics[angle=-90,clip=true]{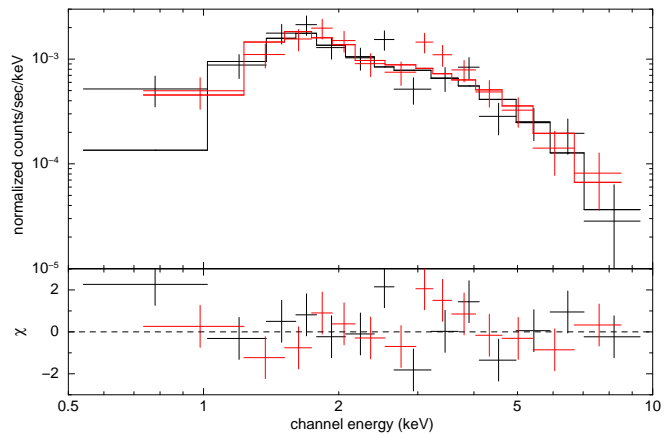}}
\caption{\textit{Top panel}: average spectrum of \saxj~with the
best--fit power--law model. The spectra of the \textit{MOS1} and
\textit{MOS2} cameras are shown in black and red, respectively.
\textit{Bottom panel}: data-model residuals, in units of
$\sigma$.}\label{spectra}
\end{figure}

A good fit was obtained with an absorbed power--law
(Fig.\ref{spectra}), yielding a hydrogen column density $N_{\rm H}
= (1.17^{+0.49}_{-0.35})\times10^{22}$ cm$^{-2}$ and a photon
index $\Gamma=1.74^{+0.43}_{-0.31}$. The absorbed flux in the
energy range 0.2--12 keV is $f_{\rm X}\sim1.45\times10^{-13}$ erg
cm$^{-2}$ s$^{-1}$, while the corresponding unabsorbed flux is
$\sim2.55\times10^{-13}$ erg cm$^{-2}$ s$^{-1}$. Acceptable fits
were also obtained with a thermal bremsstrahlung (\textit{kT} = 9.2 keV) and
with a blackbody (\textit{kT} $\sim$ 1 keV).

To study the source variability, we analyzed separately the ten
data--sets. For each of them we performed a detailed source
detection in five energy bands (0.2--0.5, 0.5--1, 1--2, 2--4.5 and
4.5--12 keV), applying the same procedure and parameters used by
the \XMM\ \SSC\ to produce the \XMM\ \textit{serendipitous source
catalogue} \citep{wat09}. The source detection was performed
simultaneously on both \MOS\ data sets and using  the
corresponding exposure maps which account for spatial quantum
efficiency variations, mirror vignetting, and effective field of
view. The threshold value for the detection likelihood was set to
5 for the \textit{likemin} parameter of the \SAS\ task {\tt
eboxdetect} and to 6 for the \textit{mlmin} parameter of the
\SAS~tasks {\tt emldetect} and \texttt{esensmap}. We emphasize
that these values imply a rather loose constraint on the source
significance in order to be detected; they allow to detect even
very weak sources, with a low (i.e. $\simeq$ 3) signal to noise
ratio. In this way we aim to check if \saxj\ is even only
marginally detected in any observation. On the other hand, in our
detection procedure we used 'ad hoc' \textit{energy conversion
factors} (\textit{ECF}) to convert the measured count rates of the
detected sources (both in each of the five energy bands and in the
total 0.2--12 keV band) into the corresponding energy flux. They
were derived based on the best--fit power--law model of the source
average spectra.

Based on this analysis, we found that \saxj\ was detected only  in
six  observations, as reported  in Table~\ref{detections}. The
upper limits on the  count rate (at the confidence level
corresponding to a detection likelihood \textit{L} = 6) are
obtained from the sensitivity maps at the source position.

\begin{table*}
\caption{Flux and luminosity values of \saxj\ in the individual
observations.
} \label{detections}
\begin{tabular}{crrrr}
 \hline
Start Observation   &   Net exposure  &   Absorbed flux  (0.2-12 keV)    &   Luminosity$^{(a)}$   & Hardness   \\
   date - UT        &     (ks)      &   (erg cm$^{-2} s^{-1}$)        &   (erg s$^{-1}$)          & Ratio$^{(b)}$  \\
\hline
2003-09-11 13:22:41   &  8.1     &   $< 6.1\times10^{-14}$       &   $< 3.03\times10^{32}$        & -   \\
2003-09-15 18:47:20   &  5.1     &   $< 7.8\times10^{-14}$       &   $< 3.88\times10^{32}$        & -   \\
2003-09-18 13:17:46   &  6.8     &   $< 2.2\times10^{-13}$       &   $< 11.1\times10^{32}$        & -   \\
2003-09-21 09:53:21   &  3.7     &   $(1.7\pm0.5)\times10^{-13}$ &   $(8.5\pm0.5)\times10^{32}$   & 0.17 $\pm$ 0.18 \\
2003-09-25 12:45:02   &  5.0     &   $(5.5\pm0.7)\times10^{-13}$ &   $(27.6\pm3.4)\times10^{32}$  & 0.48 $\pm$ 0.08 \\
2003-10-08 07:29:55   &  2.8     &   $(3.7\pm0.6)\times10^{-13}$ &   $(18.6\pm2.4)\times10^{32}$  & 0.29 $\pm$ 0.13 \\
2003-10-13 15:30:19   &  4.6     &   $(9.0\pm3.3)\times10^{-14}$ &   $(4.5\pm1.6)\times10^{32}$   & --0.28 $\pm$ 0.21 \\
2003-10-15 22:34:19   &  5.9     &   $(3.2\pm1.3)\times10^{-14}$ &   $(1.6\pm0.7)\times10^{32}$   & --0.06 $\pm$ 0.15 \\
2004-03-13 11:44:58   &  2.7     &   $(4.0\pm0.7)\times10^{-13}$ &   $(20.7\pm3.3)\times10^{32}$  & 0.30 $\pm$ 0.11 \\
2004-04-14 08:04:48   &  4.5     &   $<1.6\times10^{-13}$        &   $<8.0\times10^{32}$ & -   \\
  \hline
\end{tabular}

$^{(a)}$ In the energy range 0.2-12 keV, corrected for the
absorption, and assuming a distance of 5 kpc.

$^{(b)}$ Ratio between the source count rates in the hard (H =
2--12 keV) and soft (S = 0.2--2 keV) energy ranges, defined as
(H--S)/(H+S).
\end{table*}

\begin{figure}[h]
\centering
\resizebox{\hsize}{!}{\includegraphics[angle=-90,clip=true]{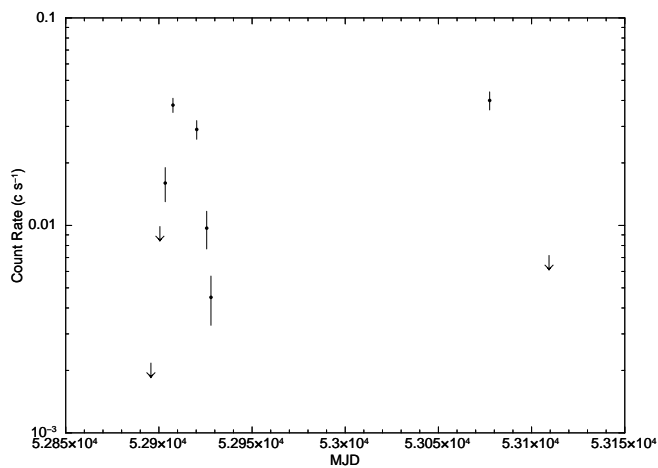}}
\caption{Light curve of \saxj\ . The count rates refer to the
0.2--12 keV and to the sum of 2 MOS. The data of the first  two
observations have been merged. The upper limits (obtained with a
threshold in  detection likelihood \textit{L} = 6) correspond to a
$\sim3\sigma$ confidence level.}
\label{luminosity}
\end{figure}

\begin{figure}[h]
\centering
\resizebox{\hsize}{!}{\includegraphics[angle=-90,clip=true]{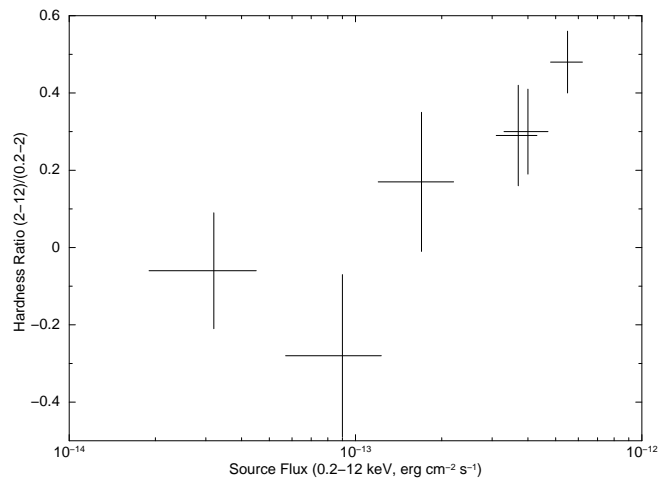}}
\caption{Hardness ratio versus source flux. The hardness ratio is
defined as (H--S)/(H+S), where H and S are the source count rates
in the hard (H = 2--12 keV) and soft (S = 0.2--2 keV) energy
ranges.} \label{HR}
\end{figure}

The long term light curve of \saxj\ is plotted in
Fig.~\ref{luminosity}. In September-October 2003 the source flux
varied by at least a factor 10. Although the observations are not
continuous, they suggest an outburst lasting about three weeks,
with a rise time of only a few days to a maximum flux of
$\sim 5 \times$10$^{-13}$ erg cm$^{-2}$ s$^{-1}$, followed by a
similarly rapid decay. A similar flux level was observed again six
months later. Comparing the hardness ratios measured in the
different observations, we found some evidence for a slight
spectral hardening correlated with the source intensity
(Table~\ref{detections}, Fig.~\ref{HR}).

\section{Discussion}

The maximum flux we observed for \saxj , 4.2 $\times$ 10$^{-13}$ erg
cm$^{-2}$ s$^{-1}$  in the 2-10 keV range, is a factor $>$ 30
smaller than that measured at the time of the BeppoSAX discovery
in 1997 (1.2 $\times$ 10$^{-11}$ erg cm$^{-2}$ s$^{-1}$,
\citet{kaa99}). To our knowledge, this is the lowest flux ever
reported for \saxj . The upper limits of the \XMM\ pointings of
September 2003 imply an even smaller flux.

The distance of \saxj\ is not well constrained. The range 2.5--5
kpc was estimated by \citet{kaa99} from the properties of the
proposed optical counterpart. We also note that a distance much in
excess 5 kpc is unlikely, considering the location of \saxj\ in
the Galactic anti-center direction. In the following discussion,
where we consider the two alternative possibilities for the origin
of the observed X--ray emission, we conservatively normalize the
relevant quantities to a distance d$_5$ of 5 kpc.

\subsection{Accretion powered X-ray emission}

Neutron stars accreting from Be star companions constitute the
large majority of the high mass X-ray binary systems present in
the Galaxy. Also in view of our improved localization of \saxj ,
it is natural to first discuss this possibility.

Already at the time of its discovery, it was noticed that the
luminosity of \saxj\ was relatively small compared to classical
Be/neutron stars systems.  The persistent sources of this class
have X--ray luminosity of 10$^{36}$--10$^{37}$ erg s$^{-1}$. This
is also the luminosity typically reached during the outbursts of
transient Be/neutron stars systems, which comprise the majority of
this population. With the advent of more sensitive observations a
number  of persistent Be binaries with lower luminosity,
$\sim$ 10$^{34}$ erg s$^{-1}$, have also been discovered (see, e.g.
\citet{rei99,lap06,lap07}). Our observations indicate for \saxj\
an average luminosity, a few 10$^{33}$ d$_{5}^{2}$ erg s$^{-1}$,
much smaller than these values,  and provide an upper limit as low
as 3$\times$10$^{32}$ d$_{5}^{2}$ erg s$^{-1}$ in mid September
2003. If we further consider that the source could well be closer
than 5 kpc, we are faced with an even smaller luminosity.

The new data clearly indicate that \saxj\ is a transient source,
but it differs from the other Be systems for its low luminosity
both during the ``high state'' and in ``quiescence''. The non
detection in September 2003 allows us to set an upper limit on the
mass accretion rate of 1.5$\times$10$^{12}$ g s$^{-1}$.  This
limit applies assuming that the accretion flow proceeds down to
the neutron star surface, which is very unlikely if indeed the
neutron star is rotating at 33.8 ms. In the  presence of the
neutron star magnetic field, different scenarios preventing, or
reducing, the accretion rate onto the neutron star surface can
occur (see, for example, \citet{cam98}). For such a short spin
period and low luminosity, the direct accretion regime, in which
the magnetospheric radius is smaller than the corotation radius
and hence the magnetic centrifugal barrier is open, requires a
magnetic field smaller than $\sim10^8$ G. This field is at least
three orders of magnitude lower  than that expected in a young
neutron star with a Be companion. In fact all other accreting
pulsars of this class have much longer spin periods. The only
exception is the recurrent transient A 0538--66 which rotates at
69 ms \citep{ski82}. Note that the pulsations in this systems were
only detected during a bright outburst reaching a luminosity of
$\sim$ 10$^{39}$ erg s$^{-1}$ \citep{ski82}, implying a magnetic
field of $\sim$ 10$^{11}$ G. Although smaller than the average,
such a field is not implausible. Furthermore it is consistent with
the interpretation of the unpulsed quiescent luminosity of A
0538--66 (several 10$^{33}$ erg s$^{-1}$)  as accretion halted at
the centrifugal barrier \citep{cam95}.

If \saxj\ has a typical magnetic field, the low luminosity
observed with XMM-Newton could be due to  mass accretion stopped
at the magnetospheric radius. Assuming for simplicity spherically
symmetric accretion, and a neutron star with mass 1.4 $\msun$ and
radius 10$^{6}$ cm, gives in this case an X-ray  luminosity of
$\sim 2\times10^{32}$ $B_{12}^{-4/7}$ $\dot{M}_{15}^{9/7}$ ergs
s$^{-1}$ (B$_{12}$ is the magnetic field in units of 10$^{12}$ G
and $\dot{M}_{15}$ the accretion rate in units of 10$^{15}$ g
s$^{-1}$; see, e.g., \citet{cam98}). However, the higher
luminosity state observed in the past with BeppoSAX and RossiXTE
cannot be explained in the same way due to the presence of
pulsations with a relatively high pulsed fraction, which are not
expected when the magnetic  centrifugal barrier operates.

In conclusion, the difficulties already pointed out in
interpreting \saxj\ as a typical accretion powered Be binary
\citep{kaa00,nic00} are reinforced by the low luminosity reported
here. Of course, the properties of this system would fit better in
this scenario if the fast periodicity were disproved by further
observations.

\subsection{Rotation powered X-ray emission}

The alternative interpretation of \saxj\ is that of a rotation
powered neutron star, whose X--ray emission derives from the shock
between the relativistic pulsar wind and the companion's wind. The
large luminosity difference between our data and the previous
observations could be due to varying shock conditions in a very
eccentric orbit. However, also in this scenario this source would
present some peculiar properties, compared to the (admittedly few)
other systems of this kind. The large variations seen in September
2003 are difficult to explain if the source was far from
periastron, where no big changes in the shock properties are
expected. Rotation powered pulsars in interacting binaries, such
as the already mentioned PSR B1259--63 \citep{cer06,cer09} or the
``black widow'' pulsar PSR B1957+20 \citep{hua07}, have X-ray
efficiencies in the range 10$^{-4}$--10$^{-2}$. On the other hand,
the $\pdot$ reported by \citep{kaa00}, corresponding to a
rotational energy loss ${\dot E}_{rot}$ larger than a few
10$^{38}$ erg s$^{-1}$, implies a much lower efficiency for \saxj
.

\section{Conclusions}

Despite \saxj\ was found by \XMM\ in a very low luminosity state,
thanks to the good sensitivity of the EPIC instrument, we could
derive a precise localization that confirms the proposed Be
optical counterpart of this source. We observed large flux
variability when the source was detected in September/October 2003
and derived a stringent upper limit of 3 $\times$ 10$^{32}$
d$_{5}^{2}$ erg s$^{-1}$ for its luminosity when the source was
not detected.

The spectral and flux properties of \saxj\ are consistent with a
variable neutron--star binary powered by accretion from the Be
companion or by the loss of rotational energy. Both
interpretations imply some peculiarities with respect to other
known sources. These result from the very short spin period and
possibly high period derivative reported for \saxj, which
unfortunately we could not confirm owing to the source faintness
during our observations.

\begin{acknowledgements}
This work is based on  observations obtained with \XMM, an ESA
science mission  with instruments  and  contributions directly
funded by  ESA Member States  and NASA.  The  \XMM~data analysis
is supported  by the Italian  Space  Agency  (ASI). The Guide Star
Catalogue II is a joint project of the Space Telescope Science
Institute and the Osservatorio Astronomico di Torino. The
Digitized Sky Surveys were produced at the Space Telescope Science
Institute under U.S. Government grant NAG W-2166.
\end{acknowledgements}

\bibliographystyle{aa}
\bibliography{pap_saxj0635_I4}

\begin{thebibliography}{19}
\expandafter\ifx\csname natexlab\endcsname\relax\def\natexlab#1{#1}\fi

\bibitem[{{Campana} {et~al.}(1998){Campana}, {Colpi}, {Mereghetti}, {Stella},
  \& {Tavani}}]{cam98}
{Campana}, S., {Colpi}, M., {Mereghetti}, S., {Stella}, L., \& {Tavani}, M.
  1998, \aapr, 8, 279

\bibitem[{{Campana} {et~al.}(1995){Campana}, {Stella}, {Mereghetti}, \&
  {Colpi}}]{cam95}
{Campana}, S., {Stella}, L., {Mereghetti}, S., \& {Colpi}, M. 1995, \aap, 297,
  385

\bibitem[{{Chernyakova} {et~al.}(2009){Chernyakova}, {Neronov}, {Aharonian},
  {Uchiyama}, \& {Takahashi}}]{cer09}
{Chernyakova}, M., {Neronov}, A., {Aharonian}, F., {Uchiyama}, Y., \&
  {Takahashi}, T. 2009, astro-ph/0905-3341

\bibitem[{{Chernyakova} {et~al.}(2006){Chernyakova}, {Neronov}, {Lutovinov},
  {Rodriguez}, \& {Johnston}}]{cer06}
{Chernyakova}, M., {Neronov}, A., {Lutovinov}, A., {Rodriguez}, J., \&
  {Johnston}, S. 2006, \mnras, 367, 1201

\bibitem[{{Cusumano} {et~al.}(2000){Cusumano}, {Maccarone}, {Nicastro},
  {Sacco}, \& {Kaaret}}]{cus00}
{Cusumano}, G., {Maccarone}, M.~C., {Nicastro}, L., {Sacco}, B., \& {Kaaret},
  P. 2000, \apjl, 528, L25

\bibitem[{{Huang} \& {Becker}(2007)}]{hua07}
{Huang}, H.~H. \& {Becker}, W. 2007, \aap, 463, L5

\bibitem[{{Johnston} {et~al.}(1992){Johnston}, {Manchester}, {Lyne}, {Bailes},
  {Kaspi}, {Qiao}, \& {D'Amico}}]{jon92}
{Johnston}, S., {Manchester}, R.~N., {Lyne}, A.~G., {et~al.} 1992, \apjl, 387,
  L37

\bibitem[{{Kaaret} {et~al.}(2000){Kaaret}, {Cusumano}, \& {Sacco}}]{kaa00}
{Kaaret}, P., {Cusumano}, G., \& {Sacco}, B. 2000, \apjl, 542, L41

\bibitem[{{Kaaret} {et~al.}(1999){Kaaret}, {Piraino}, {Halpern}, \&
  {Eracleous}}]{kaa99}
{Kaaret}, P., {Piraino}, S., {Halpern}, J., \& {Eracleous}, M. 1999, \apj, 523,
  197

\bibitem[{{La Palombara} \& {Mereghetti}(2006)}]{lap06}
{La Palombara}, N. \& {Mereghetti}, S. 2006, \aap, 455, 283

\bibitem[{{La Palombara} \& {Mereghetti}(2007)}]{lap07}
---. 2007, \aap, 474, 137

\bibitem[{{Lasker} {et~al.}(2008){Lasker}, {Lattanzi}, {McLean}, {Bucciarelli},
  {Drimmel}, {Garcia}, {Greene}, {Guglielmetti}, {Hanley}, {Hawkins},
  {Laidler}, {Loomis}, {Meakes}, {Mignani}, {Morbidelli}, {Morrison},
  {Pannunzio}, {Rosenberg}, {Sarasso}, {Smart}, {Spagna}, {Sturch},
  {Volpicelli}, {White}, {Wolfe}, \& {Zacchei}}]{Lasker2008}
{Lasker}, B.~M., {Lattanzi}, M.~G., {McLean}, B.~J., {et~al.} 2008, \aj, 136,
  735

\bibitem[{{Nicastro} {et~al.}(2000){Nicastro}, {Gaensler}, \&
  {McLaughlin}}]{nic00}
{Nicastro}, L., {Gaensler}, B.~M., \& {McLaughlin}, M.~A. 2000, \aap, 362, L5

\bibitem[{{Reig} \& {Roche}(1999)}]{rei99}
{Reig}, P. \& {Roche}, P. 1999, \mnras, 306, 100

\bibitem[{{Skinner} {et~al.}(1982){Skinner}, {Bedford}, {Elsner}, {Leahy},
  {Weisskopf}, \& {Grindlay}}]{ski82}
{Skinner}, G.~K., {Bedford}, D.~K., {Elsner}, R.~F., {et~al.} 1982, \nat, 297,
  568

\bibitem[{{Str{\"u}der} {et~al.}(2001){Str{\"u}der}, {Briel}, {Dennerl},
  {Hartmann}, {Kendziorra}, {Meidinger}, {Pfeffermann}, {Reppin}, {Aschenbach},
  {Bornemann}, {Br{\"a}uninger}, {Burkert}, {Elender}, {Freyberg}, {Haberl},
  {Hartner}, {Heuschmann}, {Hippmann}, {Kastelic}, {Kemmer}, {Kettenring},
  {Kink}, {Krause}, {M{\"u}ller}, {Oppitz}, {Pietsch}, {Popp}, {Predehl},
  {Read}, {Stephan}, {St{\"o}tter}, {Tr{\"u}mper}, {Holl}, {Kemmer}, {Soltau},
  {St{\"o}tter}, {Weber}, {Weichert}, {von Zanthier}, {Carathanassis}, {Lutz},
  {Richter}, {Solc}, {B{\"o}ttcher}, {Kuster}, {Staubert}, {Abbey}, {Holland},
  {Turner}, {Balasini}, {Bignami}, {La Palombara}, {Villa}, {Buttler},
  {Gianini}, {Lain{\'e}}, {Lumb}, \& {Dhez}}]{str01}
{Str{\"u}der}, L., {Briel}, U., {Dennerl}, K., {et~al.} 2001, \aap, 365, L18

\bibitem[{{Thompson} {et~al.}(1995){Thompson}, {Bertsch}, {Dingus}, {Esposito},
  {Etienne}, {Fichtel}, {Friedlander}, {Hartman}, {Hunter}, {Kendig}, {Mattox},
  {McDonald}, {von Montigny}, {Mukherjee}, {Ramanamurthy}, {Sreekumar},
  {Fierro}, {Lin}, {Michelson}, {Nolan}, {Shriver}, {Willis}, {Kanbach},
  {Mayer-Hasselwander}, {Merck}, {Radecke}, {Kniffen}, \& {Schneid}}]{tho95}
{Thompson}, D.~J., {Bertsch}, D.~L., {Dingus}, B.~L., {et~al.} 1995, \apjs,
  101, 259

\bibitem[{{Turner} {et~al.}(2001){Turner}, {Abbey}, {Arnaud}, {Balasini},
  {Barbera}, {Belsole}, {Bennie}, {Bernard}, {Bignami}, {Boer}, {Briel},
  {Butler}, {Cara}, {Chabaud}, {Cole}, {Collura}, {Conte}, {Cros}, {Denby},
  {Dhez}, {Di Coco}, {Dowson}, {Ferrando}, {Ghizzardi}, {Gianotti}, {Goodall},
  {Gretton}, {Griffiths}, {Hainaut}, {Hochedez}, {Holland}, {Jourdain},
  {Kendziorra}, {Lagostina}, {Laine}, {La Palombara}, {Lortholary}, {Lumb},
  {Marty}, {Molendi}, {Pigot}, {Poindron}, {Pounds}, {Reeves}, {Reppin},
  {Rothenflug}, {Salvetat}, {Sauvageot}, {Schmitt}, {Sembay}, {Short},
  {Spragg}, {Stephen}, {Str{\"u}der}, {Tiengo}, {Trifoglio}, {Tr{\"u}mper},
  {Vercellone}, {Vigroux}, {Villa}, {Ward}, {Whitehead}, \& {Zonca}}]{tur01}
{Turner}, M.~J.~L., {Abbey}, A., {Arnaud}, M., {et~al.} 2001, \aap, 365, L27

\bibitem[{{Watson} {et~al.}(2009){Watson}, {Schr{\"o}der}, {Fyfe}, {Page},
  {Lamer}, {Mateos}, {Pye}, {Sakano}, {Rosen}, {Ballet}, {Barcons}, {Barret},
  {Boller}, {Brunner}, {Brusa}, {Caccianiga}, {Carrera}, {Ceballos}, {Della
  Ceca}, {Denby}, {Denkinson}, {Dupuy}, {Farrell}, {Fraschetti}, {Freyberg},
  {Guillout}, {Hambaryan}, {Maccacaro}, {Mathiesen}, {McMahon}, {Michel},
  {Motch}, {Osborne}, {Page}, {Pakull}, {Pietsch}, {Saxton}, {Schwope},
  {Severgnini}, {Simpson}, {Sironi}, {Stewart}, {Stewart}, {Stobbart}, {Tedds},
  {Warwick}, {Webb}, {West}, {Worrall}, \& {Yuan}}]{wat09}
{Watson}, M.~G., {Schr{\"o}der}, A.~C., {Fyfe}, D., {et~al.} 2009, \aap, 493,
  339

\end{thebibliography}

\end{document}